\documentclass[prl,superscriptaddress,twocolumn,aps]{revtex4-2}
\usepackage{amssymb}
\usepackage{graphicx}
\usepackage{longtable}
\usepackage{amsmath}
\usepackage{amsfonts}
\usepackage{color}
\begin{document}

\author{A. Chiesa}
\affiliation{Dipartimento di Scienze Matematiche, Fisiche e Informatiche, Universit\`a di Parma, I-43124 Parma, Italy}
\affiliation{Institute for Advanced Simulation, Forschungszentrum J\"ulich, 52425 J\"ulich, Germany}
\author{T. Guidi}
\affiliation{ISIS facility, Rutherford Appleton Laboratory, OX11 0QX Didcot, UK}
\author{S. Carretta}
\affiliation{Dipartimento di Scienze Matematiche, Fisiche e Informatiche, Universit\`a di Parma, I-43124 Parma, Italy}
\author{S. Ansbro}
\affiliation{School of Chemistry and Photon Science Institute, The University of Manchester, M13 9PL Manchester, UK}
\affiliation{Institut Laue-Langevin, 71 Avenue des Martyrs CS 20156, Grenoble Cedex 9 F-38042, France}
\author{G. A. Timco} 
\affiliation{School of Chemistry and Photon Science Institute, The University of Manchester, M13 9PL Manchester, UK}
\author{I. Vitorica-Yrezabal}
\affiliation{School of Chemistry and Photon Science Institute, The University of Manchester, M13 9PL Manchester, UK}
\author{E.Garlatti}
\affiliation{Dipartimento di Scienze Matematiche, Fisiche e Informatiche, Universit\`a di Parma, I-43124 Parma, Italy}
\author{G. Amoretti}
\affiliation{Dipartimento di Scienze Matematiche, Fisiche e Informatiche, Universit\`a di Parma, I-43124 Parma, Italy}
\author{R. E. P. Winpenny}
\affiliation{School of Chemistry and Photon Science Institute, The University of Manchester, M13 9PL Manchester, UK}
\author{P. Santini}
\affiliation{Dipartimento di Scienze Matematiche, Fisiche e Informatiche, Universit\`a di Parma, I-43124 Parma, Italy}

\date{\today }
\title{Magnetic exchange interactions in the molecular nanomagnet Mn$_{12}$}

\begin{abstract}
The discovery of magnetic bistability in Mn$_{12}$ more than 20 years ago marked the birth of molecular magnetism, an extremely fertile interdisciplinary field and a powerful route to create tailored magnetic nanostructures. However, the difficulty to determine the interactions within the core of complex polycentric molecules often prevents their understanding and can hamper addressing important fundamental and applicative issues. Mn$_{12}$ is an outstanding example: although it is the forefather and most studied of all molecular nanomagnets, an unambiguous determination even of the leading magnetic exchange interactions is still lacking. Here we exploit four-dimensional inelastic neutron scattering to portray how individual spins fluctuate around the magnetic ground state, thus fixing the exchange couplings of Mn$_{12}$ for the first time. Our results demonstrate the power of four-dimensional inelastic neutron scattering as an unrivalled tool to characterize magnetic clusters.
\end{abstract}
% \pacs{75.50.Xx, COMPLETE}
\maketitle

\vskip 3 cm

\begin{figure}
	\centering
		\includegraphics[width=8 cm]{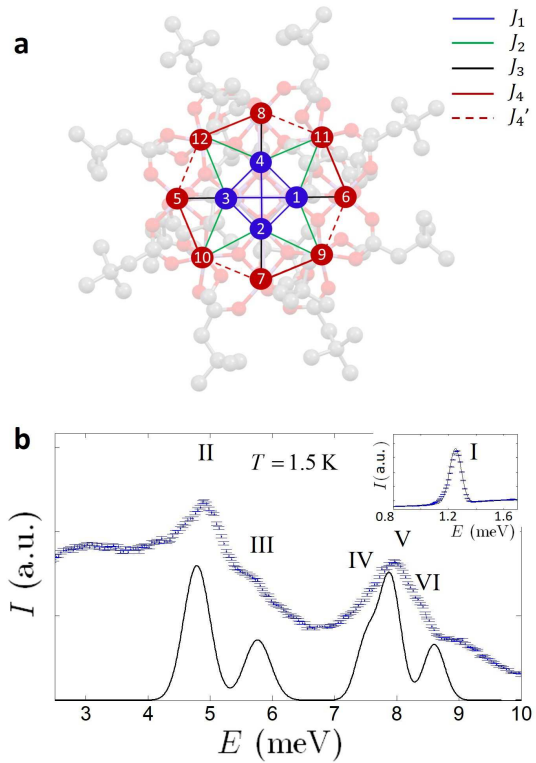}
	\caption{(a) Scheme of the Mn$_{12}$-$^t$BuAc molecule, with different lines representing the relevant different exchange interactions. Red circles: Mn$^{3+}$ ions ($s=2$). Blue circles: Mn$^{4+}$  ions ($s=3/2$). Seven distinct exchange constant are allowed by the $S_4$ symmetry of the molecule, but most models assume only four parameters ($J_{1-4}$ with $J_4' = J_4$) because of similarities in some exchange paths. (b) INS spectrum collected on LET at 1.5 K, using 15.4 meV (and 4.2 in the inset) incident neutron energy. The continuous line is the corresponding simulation with the best-fit parameters (in meV) $J_1$ = -1.2, $J_2$  = 3.2, $J_3$ = 6.6, $J_4$ =0.55, $J_4'$ = 0.30, $d$ = -0.315. Eigenstates are listed in Table S1 \cite{SUPPL}. Peak \textbf{I} corresponds to a transition between states $M = \pm 10$ and $M = \pm 9$ of the ground $S=10$ multiplet. The slight asymmetry of the peak is due to the instrumental resolution function of the TOF spectrometer. Peaks \textbf{II}-\textbf{VI} are inter-multiplet transitions to different excited $S=9$ multiplets. The broad peak at 3 meV is a phonon, as shown by the monotonic increase as $Q^2$ of the associated form factor. }
	%(c)  $\chi T$ vs. $T$ (inset: $\chi$  vs. $T$). Calculations with the best-fit parameters (lines) are in very good agreement with experiments \cite{26} [dots].}
	\label{fig1}
\end{figure}

\begin{figure}
	\centering
		\includegraphics[width=8 cm]{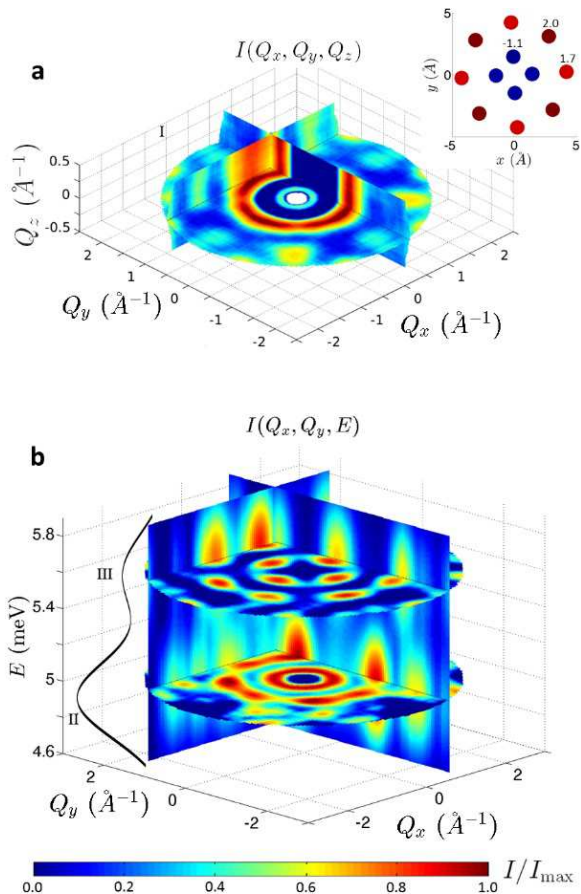}
	\caption{(a) Form factor for the intra-multiplet transition $|S=10, M=\pm 10\rangle \rightarrow |S=10, M=\pm 9\rangle$, i.e., $S(E,{\bf Q})$ for $E$ = 1.25 meV (giant-spin excitation \textbf{I}). The inset shows the equivalent real-space information, that is the distribution of the static magnetization of the giant spin over the three inequivalent Mn sites, $\mu_n = \langle S=10, M= 10|s_{z n} |S=10, M= 10\rangle$. The values (in  $\mu_{\rm B}$)  $\mu_1$=-1.2 (2),  $\mu_6$ = 1.7 (0.15) and  $\mu_{11}$ =2 (0.15) are extracted directly from the form-factor and compare well with polarized neutron diffraction (-1.17, 1.84, 1.90) \cite{28} and NMR (-1.3, 1.8, 1.8) \cite{29} data on a slightly different variant of Mn$_{12}$. (b) $S(E,{\bf Q})$ as a function of $Q_x$, $Q_y$ and $E$, and integrated over the full $Q_z$ range \cite{footnote}. The energy window contains the intermultiplet peaks \textbf{II} and \textbf{III} of Fig. \ref{fig1}b.}
	\label{fig2}
\end{figure}

\begin{figure*}
	\centering
		\includegraphics[width=16cm]{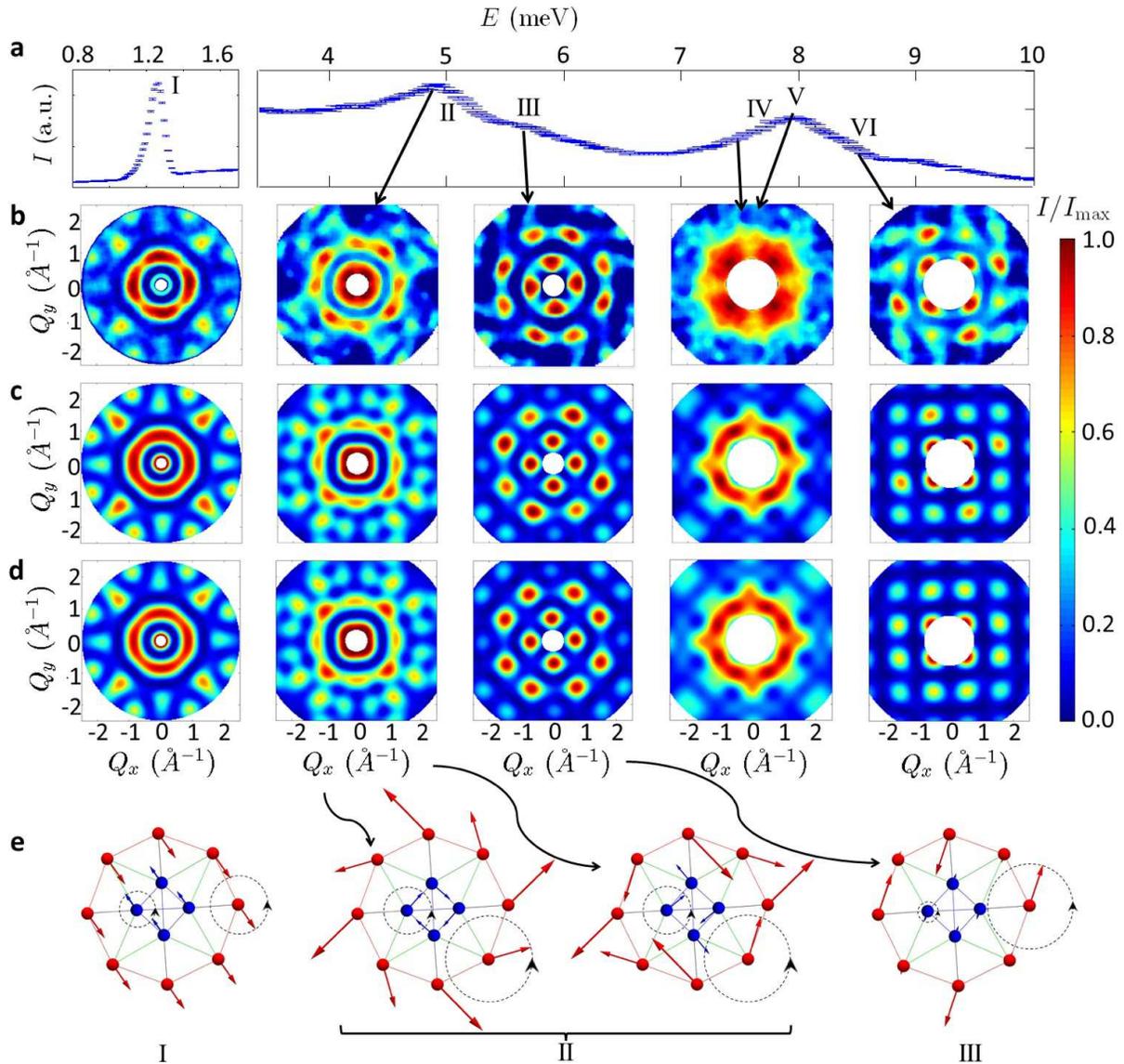}
	\caption{(a) INS energy spectrum (same as Fig.1b). (b) Constant-energy cuts for $S(E,{\bf Q})$, integrated over the full $Q_z$ range, obtained from measurements at $T=$1.5 K for incident neutron energies of 4.2 meV (first column, peak \textbf{I}) and 15.4 meV (peaks \textbf{II}, \textbf{III}, \textbf{IV}, \textbf{V}, \textbf{VI}) \cite{footnote}.  Each map is normalized to its maximum. (c) Corresponding simulated maps, obtained with parameters (in meV)  $J_1$ = -1.2 (1), $J_2$ = 3.2 (2), $J_3$ = 6.6 (3), $J_4$ = 0.55 (5), $J_4'$ = 0.30 (5) , $d$=-0.315 (2). Eigenstates are listed in Table S1 \cite{SUPPL}. Row (d) highlights the effect of a slight variation of exchange parameters ($J_4 = J_4'$=0.42 meV is assumed).  Peaks \textbf{IV} and \textbf{V} are too close in energy to extract individual maps, and only their sum is addressed. (e) Precession pattern of the individual Mn spins for excitations \textbf{I}, \textbf{II} and \textbf{III}. For each excitation, arrows represent the twelve vectors $(\langle s_{x n}(t), s_{y n}(t) \rangle)$  describing the spatial pattern of the spins preceding around $z$, after a resonant perturbation has brought a molecule from its $M=10$ ground state into a superposition state with a small component on the corresponding excited $M=9$ state. All the spins precede with the same frequency $E/h$ and dashed circular arrows indicate the direction of the spin precessions for two representative sites. Preparing the system in an initial state with opposite $M$ would induce an opposite precession of the spins. The two panels for excitation \textbf{II} correspond to a pair of degenerate states (Table S1 \cite{SUPPL}). For peaks \textbf{IV}, \textbf{V} and \textbf{VI} experimental form factors are more noisy or unresolved. Their precession pattern is not directly deduced from data, and is obtained by simulations of the best-fit Hamiltonian (Fig. S5 in \cite{SUPPL}).}
	\label{fig3}
\end{figure*}

The ability to store magnetic information in a single molecule was reported for the first time in the Mn$_{12}$ polymetallic complex \cite{1}. Many further breakthroughs followed from studies of this molecule, including the observation of macroscopic quantum tunneling of magnetization \cite{2,Luis}, and the discovery that it can be used to build devices based on the Grover algorithm \cite{3}. The phrase ''single molecule magnet" was invented to describe the physics of Mn$_{12}$, and this molecule inspired the entire field of molecular magnetism, which continues to produce remarkable science \cite{4,5,6,7,8,9,10,11,12,13,14,15,16,17}.  However, the understanding of complex polycentric molecules is often limited due to the difficulty to determine the interactions within the core, thus hampering the addressing of important fundamental and applicative issues. Mn$_{12}$ is a particularly striking example: in spite of hundreds of papers there is not even an unambiguous description for the leading interactions of this archetypal molecule, twenty-five years after it fathered a new field of science. Thus, the debate about Mn$_{12}$ is still completely open, as witnessed by recent studies \cite{22,23,24}.\\
The phenomenology of molecular nanomagnets results from a number of interactions in the magnetic core, where isotropic exchange couplings are usually leading and various types of anisotropic single- and two-ion terms act perturbatively. The interplay of these interactions results in a multitude of physical behaviors, usually described in terms of simplified effective models. These are parametrized to capture distinctive low-temperature and low-frequency properties, but in many cases with complex cores the determination of the fundamental underlying spin Hamiltonian is still a challenge. Molecules displaying magnetic remanence like Mn$_{12}$ are usually described in terms of phenomenological ''giant spin" models, where a single quantum spin $S$ ($S=10$ in Mn$_{12}$) represents the magnetic core as a whole \cite{18,19}. Although this approach is cost-effective in terms of model complexity, it leaves in the shadows the nature of the giant spin at the atomic level, hindering the tailoring of the magnetic core for improved performance in fundamental or applicative issues \cite{20}. Moreover, the many-spin character of the core emerges already in the low-energy physics (see, e.g., \cite{10,21}).
Here we close this long-standing unresolved case: we exploit four-dimensional inelastic neutron scattering \cite{7} to portray the spin precession patterns, which are unambiguous fingerprints of the magnetic Hamiltonian, and we thus pinpoint the exchange couplings of Mn$_{12}$ for the first time. Our results open unprecedented prospects in understanding magnetic spin clusters and motivate the synthesis of new polycentric nanostructures, where the set of interactions is optimal for specific fundamental issues or applications.\\
Most of the proposed models for Mn$_{12}$ are based on a set of four isotropic exchange parameters, reported in the schematic representation in Fig. \ref{fig1}a (with $J_4 = J_4'$). The spin Hamiltonian also includes anisotropic terms accounting for the uniaxial behavior of the system and it reads:
\begin{equation}\label{hamiltonian}	
H=\sum_{m<n} J_{m n} {\bf s}_m\cdot {\bf s}_n + d \sum_{m=5}^{12} s_{z m}^2 ,
\end{equation}
where the pairs of ions included in the first sum are indicated in Fig. \ref{fig1}a and zero-field splitting terms are considered only on the eight highly-anisotropic Mn$^{3+}$ ions. We have checked that more complex choices for the anisotropic term (e.g., small non-axial terms or higher-order contributions) do not significantly affect the determination of exchange constants and are here neglected for simplicity \cite{SUPPL}.
As stated above, a firm set of parameters for Eq. \ref{hamiltonian} could not be found yet. Indeed, the known excitation energies \cite{25} only provide a coarse characterization of the spin Hamiltonian through its eigenvalues. In particular, they lack the selective information associated with the structure of eigenvectors, i.e., how individual atomic spins move when excitations are triggered. Here we use the four-dimensional inelastic neutron scattering (4D-INS) technique \cite{7} to extract such information for the low-energy transitions shown in Fig. \ref{fig1}b. The power of the technique comes from the capability to measure the scattering cross-section $S(E, \bf{Q})$ over large portions of the energy-wavevector $(E, {\bf Q})$ space, yielding a faithful portray of spin fluctuations on the space- and time-scales characterizing the internal dynamics of the magnetic core. This experimental information fingerprints the eigenstates of the spin Hamiltonian, thus enabling us to fix the value of exchange couplings univocally for the first time.
We have studied specifically Mn$_{12}-^t$BuAc (full formula [Mn$_{12}$O$_{12}$(O$_2$CCD$_2$C(CD$_3$)$_3$)$_{16}$(CD$_3$OD)$_4$]$\bullet$(C$_2$H$_5$)$_2$O) which crystallizes with $S_4$ symmetry, and it is the deuterated analog of the isostructural [Mn$_{12}$O$_{12}$(O$_2$CCH$_2$Bu$^t$)$_{16}$(MeOH)$_4$]$\bullet$MeOH molecule \cite{26} \cite{SUPPL}.
%Single crystals large enough for 4D-INS studies were grown by slow evaporation, under N$_2$ at ambient temperature, of a solution of the Mn$_{12}$-tBuAc compound dissolved in a mixture of diethyl ether/methanol-d4 in presence of a small amount of deuterated t-butylacetic acid.
Measurements have been performed on the high-resolution LET spectrometer at ISIS \cite{LET}, on a collection of oriented single crystals \cite{SUPPL}. Figure \ref{fig1}b shows INS spectra taken at $T = 1.5$ K at two different incident neutron wavelengths, with five peaks clearly distinguishable between 1 and 10 meV. Previous INS studies of related Mn$_{12}$ molecules \cite{25} have assigned these peaks to transitions from the ground spin doublet $|S=10, M=\pm 10\rangle$ ($S$ is the total-spin quantum number), as the population of any other state is negligible at this temperature. For example, the lowest-energy peak (labeled \textbf{I}) at 1.25 meV represents the intra-multiplet transition to the $|S=10, M=\pm 9\rangle$ doublet.
Using the 4D-INS technique we can obtain far more information (Figure \ref{fig2}). 4D-INS data reported in Figure \ref{fig2}a directly demonstrate (see the discussion of Figure \ref{fig3}e below) that excitation \textbf{I} corresponds to a precession of the giant spin around the anisotropy axis, with its internal structure kept rigid. As discussed in \cite{27} and in \cite{SUPPL}, this information is equivalent to that contained in the distribution of the giant-spin moment over different Mn ions, i.e., the set of expectation values
$\pm \mu_n = \langle S=10, M=\pm 10|s_{z n} |S=10, M=\pm 10\rangle$.
The measured form-factor (i.e., the ${\bf Q}$-dependence of the scattering intensity) of Fig. \ref{fig2}a enables us to extract the set of $\mu_n$ values in the inset, which are in line with those determined by neutron diffraction \cite{28} and NMR \cite{29}. The moment distribution reveals ferromagnetic correlations among the four Mn$^{4+}$ and among the eight Mn$^{3+}$ spins, with the two sets antiferromagnetically correlated to each other. However, the Mn$^{3+}$ and Mn$^{4+}$ moments are significantly below saturation,indicating that the spins are not locked in a maximally-aligned state due to quantum fluctuations (see Table S1 in \cite{SUPPL}).\\
The local distribution of moments, $\mu_n$, is stable over a range of exchange constants and hence is not sufficient to fix the magnitude of the exchange interactions $J_{m n}$ uniquely. It is intuitively clear that exchange is probed more effectively through excitations that break the internal alignment of Mn spins in their ground state. Just like spin waves in bulk magnetic compounds, their energies and structure directly reflect the values of exchange constants. These excitations correspond to peaks {\bf II}-{\bf VI} in Fig. \ref{fig1}b and represent inter-multiplet transitions between the ground $|S=10, M=\pm 10\rangle$ doublet and a set of excited   $|S=9, M=\pm 9\rangle$ doublets. Although these ${\bf Q}$-integrated energy spectra, together with susceptibility (Fig. S2 in \cite{SUPPL}), provide constraints on the set of exchange constants, they are not selective. Conversely, a clear identification of the $|S=9\rangle$  wavefunctions is achieved thanks to the measured ${\bf Q}$-dependencies, which contain detailed information on the composition and symmetry of the states involved in the transition.  For example, Fig. \ref{fig2}b shows $S(E,{\bf Q})$ as a function of $Q_x$, $Q_y$ and $E$, and integrated over the full $Q_z$ range. The energy interval spans peaks {\bf II} and {\bf III} of Fig. \ref{fig1}b, whereas constant-energy cuts of $S(E,{\bf Q})$ for $E$ corresponding to all the peaks in Fig. \ref{fig1}b are shown in Fig.\ref{fig3}. The great amount of information available in these experimental data is immediately evident.
The $S(E,{\bf Q})$ data fully characterize the low-lying multiplets, and make it possible to identify the five exchange parameters. The simulation of these data (Figure \ref{fig3}c) unequivocally establishes the five exchange parameters (in meV): $J_1$ = -1.2(1), $J_2$ = 3.2(2), $J_3$ = 6.6(3), $J_4$ = 0.55(5), $J_4'$ = 0.30(5). The agreement between calculation and experiment is very good and the model also fits the magnetic susceptibility (Fig. S2 in \cite{SUPPL}) and peak positions (Figure \ref{fig1}b). There is just a slight discrepancy for the position of peak {\bf VI}, whose fine-tuning requires additional small parameters in Eq. \ref{hamiltonian} \cite{SUPPL}. As expected in broad terms from the internal structure of the giant-spin (Fig. \ref{fig2}a), antiferromagnetic couplings between Mn$^{3+}$ and Mn$^{4+}$ ions are leading. The coupling between the four Mn$^{4+}$ ions is ferromagnetic, whereas that between the eight Mn$^{3+}$ ions is weakly antiferromagnetic.\\
The information on eigenstates is so rich that even subtle variations of exchange parameters alter these maps. For instance, a single parameter is usually assumed for the external Mn$^{3+}$ ring, i.e., $J_4 = J_4'$ in Fig. \ref{fig1}a. Although these constants are an order of magnitude smaller than the leading ones, by enabling $J_4 \neq J_4'$ we can quantify them separately \cite{SUPPL}. The effect of the difference $J_4 - J_4'$ stands out in Fig. \ref{fig3}, showing also (panel d) simulations obtained with $J_4 = J_4' = 0.42$ meV. The intensity distribution in the intermultiplet maps is noticeably different, reflecting a change in composition of the excited $S=9$ multiplets. It is worth noting that small model variations of this type have significant impact on these maps, but negligible effects on the energy spectrum and susceptibility.\\
The information on eigenstates collected in reciprocal $(E, {\bf Q})$ space  can be made intuitive by using an equivalent description in terms of time and position variables, i.e., by portraying the precession pattern of the twelve Mn spins associated with each excitation. Indeed, the {\bf Q} dependence of a peak at energy $E_p$ reflects the spatial pattern of the spins preceding around $z$ with frequency $E_p/h$, after a resonant perturbation has brought a molecule from its $M=10$ ground state into a superposition state with a small component on the corresponding excited $M=9$ state \cite{SUPPL}. These precession motions are in a one to one correspondence with the form factor $S(E_p,{\bf Q})$, as both are set by the same reduced matrix elements \cite{SUPPL}. For a generic weak perturbation (e.g., a  $\delta$-pulse), the resulting motion will then be a weighted superposition of these single-frequency contributions.\\
Precession patterns, directly extracted from experimental data, are shown in Fig. \ref{fig3}e and represent the molecular counterpart of spin-wave excitations in bulk ferromagnets. The difference in the spin dynamics associated to the various transitions is evident: in transition \textbf{I} all the spins rigidly precess conserving the same total-spin modulus of the ground state, as expected for a giant spin excitation. Conversely, the precession pattern of all the other peaks is characterized by a zero total spin, demonstrating the inter-multiplet nature of the transitions. In addition, the different symmetries of the excited states (Table S1, \cite{SUPPL}) produce clear signatures in the precession patterns.\\
The present results finally characterize the exchange interactions in the archetypal single-molecule magnet Mn$_{12}$, enabling us to draw for the first time a sound picture of the eigenstates beyond the giant spin model. This will be the starting point to address important issues in the understanding of this molecule, which are still not really solved after more than twenty years of research. For instance, the relaxation dynamics of Mn$_{12}$ should be influenced by the low-lying excited multiplets, partially overlapping with the ground one (e.g., these lead to additional relaxation and tunneling pathways with respect to the giant spin model). In general, these results open remarkable perspectives in understanding nanomagnets with complex polycentric core. These are still relatively little explored and understood but are of fundamental importance, with potential applications in the longer term. We mention among others molecules where the role of anisotropy is not perturbative, like in presence of Co \cite{30} or $f$-electron ions \cite{17}. These can convey their large anisotropy through exchange to the whole core, thus producing large anisotropy barriers or exotic magnetic states (e.g., toroidal or chiral). On the opposite side, we mention small-anisotropy molecules where the pattern of exchange couplings results in frustration, which is important both for fundamental and applicative issues \cite{15}. More generally, experiments such as the present one show that 4D-INS is an unrivalled tool for characterizing magnetic clusters where the size and complexity of the spin structure make impossible or ambiguous the interpretation by more conventional routes.\\
We gratefully acknowledge Dr. Peixun Li for the preparation of deuterated products. A.C., S.C., E.G., G.A. and P.S acknowledge financial support from the FIRB Project No. RBFR12RPD1 and PRIN Project 2015 No. HYFSRT of the MIUR (I). A. C. acknowledges "Fondazione Angelo Della Riccia" for financial support. G.T., I.J.V.-Y. and R.E.P.W. thank the EPSRC(UK) for support, including funding for an X-ray diffractometer (grant number EP/K039547/1).


\begin{thebibliography}{99}

\bibitem{1} R. Sessoli, D. Gatteschi, A. Caneschi and M. A. Novak, Nature \textbf{365}, 141 (1993).
\bibitem{2} L. Thomas, F. Lionti, R. Ballou, D. Gatteschi, R. Sessoli and B. Barbara, Nature \textbf{383}, 145 (1996).
\bibitem{Luis} J. M. Hern\'andez, X. X. Zhang, F. Luis, J. Bartolom\'e, J. Tejada and R. Ziolo, Europhys. Lett. \textbf{35}, 301 (1996).
\bibitem{3} M. N. Leuenberger and D. Loss, Nature \textbf{410}, 789 (2001).
\bibitem{4} S. Thiele, F. Balestro, R. Ballou, S. Klyatskaya, M. Ruben and W. Wernsdorfer, Science \textbf{344}, 1135 (2014).
\bibitem{5} M. Shiddiq, D. Komijani, Y. Duan, A. Gaita-Ari\~{n}o, E. Coronado and S. Hill, Nature \textbf{531}, 348 (2016).
\bibitem{6} C. Cervetti, A. Rettori, M.G. Pini, A. Cornia, A. Repolles, F. Luis, M. Dressel, S. Rauschenbach, K. Kern, M. Burghard and L. Bogani, Nature Mater. \textbf{15}, 164 (2016).
\bibitem{7} M. L. Baker, T. Guidi, S. Carretta, J. Ollivier, H. Mutka, H. U. G\"{u}del, G. A. Timco, E. J. L. McInnes, G. Amoretti,  R. E. P. Winpenny and P. Santini,  Nature Phys. \textbf{8}, 906 (2012).
\bibitem{8} W. Wernsdorfer and R. Sessoli, Science \textbf{284}, 133 (1999).
\bibitem{9} W. Wernsdorfer, M. Murugesu and G. Christou, Phys. Rev. Lett. \textbf{96}, 057208 (2006).
\bibitem{10} S. Carretta, T. Guidi, P. Santini, G. Amoretti, O. Pieper, B. Lake, J. van Slageren, F. El Hallak, W. Wernsdorfer, H. Mutka, M. Russina, C. J. Milios and E. K. Brechin, Phys. Rev. Lett. \textbf{100}, 157203 (2008).
\bibitem{11} A. Furrer and O. Waldmann, Rev. Mod. Phys. \textbf{85}, 367 (2013).
\bibitem{12} J. Luzon, K.Bernot, I. J. Hewitt, C. E. Anson, A. K. Powell and R. Sessoli, Phys. Rev. Lett. \textbf{100}, 247205 (2008).
\bibitem{13} L. Ungur, S. K. Langley, T. N. Hooper, B. Moubaraki, E. K. Brechin, K. S. Murray and L. Chibotaru, J. Am. Chem. Soc. \textbf{134}, 18554 (2012).
\bibitem{14} E. Garlatti, T. Guidi, S. Ansbro, P. Santini, G. Amoretti, J. Ollivier, H. Mutka, G. A. Timco, I. Vitorica-Yrezabal, G. Whitehead, R. E. P. Winpenny and S. Carretta, Nature Commun. \textbf{8}, 14543 (2017).
\bibitem{15} C. Schr\"{o}der, H.-J. Schmidt, J. Schnack, and M. Luban, Phys. Rev. Lett. \textbf{94}, 207203 (2005).
\bibitem{16} S. Carretta, P. Santini, G. Amoretti, T. Guidi, J. R. Copley, Y. Qiu, R. Caciuffo, G. A. Timco and R. E. P. Winpenny, Phys. Rev. Lett. \textbf{98}, 167401 (2007).
\bibitem{17} M. S. Dutkiewicz, J. H. Farnaby, C. Apostolidis, E. Colineau, O. Walter, N. Magnani, M. Gardiner, J. B. Love, N. Kaltsoyannis, R. Caciuffo and P. L. Arnold, Nature Chem. \textbf{8}, 797-802 (2016).
\bibitem{22} V. Mazurenko, Y. O. Kvashnin, F. Jin, H. A. De Raedt, A. I. Lichtenstein and M. I. Katsnelson, Phys. Rev. B \textbf{89}, 214422 (2014).
\bibitem{23} S. G. Tabrizi, A. V. Arbuznikov and M. Kaupp, J. Phys. Chem. A \textbf{120}, 6864 (2016).
\bibitem{24} O. Hanebaum and J. Schnack, Phys. Rev. B \textbf{92}, 064424 (2015).
\bibitem{18} R. Caciuffo, G. Amoretti, A. Murani, R. Sessoli, A. Caneschi and D. Gatteschi, Phys. Rev. Lett. \textbf{81}, 4744 (1998).
\bibitem{19} I. Mirebeau, M. Hennion, H. Casalta, H. Andres, H. U. G\"{u}del, A. V. Irodova, and A. Caneschi, Phys. Rev. Lett. \textbf{83}, 628 (1999).
\bibitem{20} Molecular Magnetic Materials: Concepts and Applications (eds B. Sieklucka, and D. Pinkowicz) (Wiley-VCH, 2017).
\bibitem{21} A.-L. Barra, A. Caneschi, A. Cornia, D. Gatteschi, L. Gorini, L.-P. Heiniger, R. Sessoli and L. Sorace,  J. Am. Chem. Soc. \textbf{129}, 10754 (2007).
\bibitem{SUPPL} See Supplemental Material at [URL will be inserted by publisher] for details on synthesis, crystallography, neutron scattering experiments and calculations.
\bibitem{25} G. Chaboussant, A. Sieber, S. Ochsenbein,  H.-U. G\"udel, M. Murrie, A. Honecker, N. Fukushima, and B. Normand, Phys. Rev. B \textbf{70}, 104422 (2004).
\bibitem{26} C. Lampropoulos, M. Murugesu, A. G. Harter, W. Wernsdofer, S. Hill, N. S. Dalal, A. P. Reyes, P. L. Kuhns, K. A. Abboud and G. Christou, Inorg. Chem. \textbf{52}, 258 (2013).
\bibitem{LET} R. I. Bewley, J. W. Taylor abd S. M. Bennington, Nuclear Instruments and Methods in Physics \textbf{637}, 128 (2011).
\bibitem{footnote} We have exploited the fourfold symmetry of the maps and summed together the cross-sections obtained for wavevectors equivalent by rotations, in order to improve data statistics (see also Supplemental Material).
\bibitem{27} O. Waldmann, R. Bircher, G. Carver, A. Sieber, H. U. G\"{u}del and H. Mutka, Phys. Rev. B \textbf{75}, 174438 (2007).
\bibitem{28} R. A. Robinson, P. J. Brownd, D. N. Argyriou D. N. Hendrickson and S. M. J. Aubin, J. Phys.: Condens. Matter. \textbf{12}, 2805 (2000).
\bibitem{29} Y. Furukawa, K. Watanabe, K. Kumagai, F. Borsa and D. Gatteschi, Phys. Rev. B \textbf{64}, 104401 (2001).
\bibitem{30} M. Murrie, Chem. Soc. Rev. \textbf{39}, 1986 (2010).
%\bibitem{31} G. M. Sheldrick, Acta Crystallogr. C\textbf{71}, 3 (2015).
%\bibitem{32} Dolomanov, O. V., Bourhis, L. J., Gildea, R. J., Howard, J. A. K., Puschmann, H. OLEX2: a complete structure solution, refinement and analysis program, J. Appl. Cryst. 42, 339-341 (2009).
%34. Ewings, R. A., Buts, A., Lee, M. D., van Duijn, J., Bustinduy, I. & Perring, T. G. Horace: Software for the analysis of data from single crystal spectroscopy experiments at time-of-flight neutron instruments, Nuclear Instruments and Methods in Physics Research Section A: Accelerators, Spectrometers, Detectors and Associated Equipment 834, 132-142 (2016).





\end{thebibliography}
\end{document}